\documentclass[aps,prd nofootinbib, twocolumn,groupedaddress]{revtex4}

\usepackage{graphicx}  
\usepackage{dcolumn}   
\usepackage{bm}        
\usepackage{amssymb}   
\usepackage{amsmath}
\usepackage{amsfonts}
\usepackage{dsfont}

\def\njs{$n_{jet}^*$}
\def\co{coannihilation~}
\def\pts{$P_T^{\rm * miss}$}
\def\pt{$P_T^{\rm miss}$}

\def \cha{\widetilde{\chi}^{\pm}_1}

\def \na{\widetilde{\chi}^{0}_1}
\def \nb{\widetilde{\chi}^{0}_2}
\def \nc{\widetilde{\chi}^{0}_3}

\def \g{\widetilde{g}}
\def \ql{\widetilde{q}_L}
\def \qr{\widetilde{q}_R}

\def \ta{\widetilde{t}_1}

\def \ba{\widetilde{b}_1}

\def \sta{\widetilde{\tau}_1}

\def \slr{\widetilde{l}_R}

\def \snl{\widetilde{\nu}_{\tau}}

\def\beq{\begin{equation}}
\def\be{\begin{equation}}
\def\beqn{\begin{eqnarray}}
\def\ee{\end{equation}}
\def\eeq{\end{equation}}
\def\eeqn{\end{eqnarray}}

\begin{document}

\title{Decoding the Mechanism for the Origin of Dark Matter \\ in the Early Universe Using LHC Data  }
\author{Daniel Feldman, Zuowei Liu\footnote{Address after Sept. 1, 2008: C. N. Yang Institute for Theoretical Physics,
Stony Brook University,  Stony Brook, NY 11794, USA.}, and Pran Nath}
 \affiliation{Department of Physics,
Northeastern University, Boston, MA 02115, USA. }

\begin{abstract}
It is shown that LHC data  can allow one to decode the mechanism by which dark matter is
generated in the early universe in supersymmetric theories. We focus on two
of the major mechanisms for such
generation of dark matter  which are known to be the Stau Coannihilation (Stau-Co)  where the neutralino is typically
Bino like and annihilation on the Hyperbolic Branch (HB)  where the neutralino   has a significant Higgsino
component. An investigation of how one may discriminate between the Stau-Co region
and the HB region using LHC data is given for the mSUGRA model. The analysis utilizes
several signatures including  multi leptons, hadronic jets, b-tagging, and missing transverse momentum.
A study of the SUSY signatures reveals several correlated smoking gun signals allowing
a clear discrimination between the Stau-Co and the HB regions where dark matter  in the early
universe can originate.
\end{abstract}

\maketitle
\emph{Introduction}.
In the near future, data from the LHC
will be available allowing one to test models of physics
beyond the Standard Model (SM).
Supersymmetry (SUSY) and more specifically supergravity grand
unified models\cite{msugra,hlw,review} provide a well motivated framework for
 the exploration of new physics.
Thus  supergravity  grand unified models
 naturally lead to the lightest
neutralino as the lightest SUSY particle, or the LSP, over a significant part  of the parameter
space and with R parity it is then a  candidate for dark
matter.
  An analysis of the relic density of the LSP reveals three broad regions where the WMAP \cite{Spergel:2006hy} constraints
  are satisfied: these include  (1) the Hyperbolic Branch (HB)\cite{hb,dmdd}
   where
  multi TeV scalars can appear consistent with small fine tuning (this region is alternately referred
  sometimes as the Focus Point region (FP) or as HB/FP),  (2) the
  coannihilation regions\cite{Griest:1990kh,Drees:1992am,Mizuta:1992qp,Ellis,Nihei:2002sc},
   (3) the Higgs pole region\cite{Nath:1992ty}
   (for recent work on these regions(1-3) see \cite{Arnowitt, Feldman:2007zn, Baer:2007ya, Feldman:2008hs}).
   Of these, the stau \co region  and the HB region   are more generic while the
  pole  region (light Higgs and the CP odd Higgs $A$)  is more fine tuned.
     In addition there is also the parameter space in the bulk region
  where the relic density is satisfied due to a  combination of effects.
  An interesting issue relates to the following: to what extent the LHC data will allow one to
decode the mechanism by which dark matter is generated in the early universe.
Specifically we will focus on
dark matter originating in the  Stau-Co region or in the HB region to answer this question.

For concreteness
 we work within the framework of the mSUGRA model\cite{msugra}, which is characterized
by the parameters $m_0, m_{1/2}, A_0, \tan\beta$ and sign$(\mu)$, where $m_0$ is the
universal scalar mass, $m_{1/2}$ is the universal gaugino mass,
$A_0$ is the universal trilinear coupling
(all at the grand unification scale), and $\tan\beta \equiv \langle H_2\rangle/\langle H_1\rangle$
 where $H_2$ gives mass to  the up quarks and $H_1$ gives mass to the
 down quarks and the leptons, while $\mu$ is the Higgs mixing parameter.
 The parameter space of mSUGRA investigated in this work is  as in \cite{Feldman:2007zn}. 
In this framework we will
show that the LHC signatures carry sufficient  information for the discrimination
between the Stau-Co and the HB regions.

In the analysis, the sparticle masses and mixings are derived from
the GUT scale with the SuSpect code \cite{Djouadi:2002ze}  coupled
to micrOMEGAs \cite{Belanger:2007zz}. We merge the models via the
SUSY Les Houches Accord format \cite{SLHA} into Pythia
\cite{Sjostrand:2006za} for the computation  of SUSY production at the
LHC, in concert with PGS4 \cite{PGS4}, to simulate LHC detector
effects, and obtain the final event record. The models are
constrained by their ability to properly break electro-weak symmety,
by sparticle mass limits from LEP and Tevatron analyses, FCNC
constraints including $b \to s \gamma$ and $B_{s} \to
\mu^{+} \mu^{-}$, by the supersymmetric contribution to the muon
anomalous magnetic moment, and the double sided
bound on the relic density (see \cite{Djouadi:2006be,Feldman:2008hs}
for details). Our post trigger level cuts are as given in
\cite{Feldman:2008hs} and are standard and  we list them below:
(1) In an event, we only select photons, electrons, and muons that
have transverse momentum $P_T(p)>10$ GeV and $|\eta(p)|<2.4$,
$p=(\gamma,e,\mu)$;  (2) For hadronically decaying tau (jets):
$P_T(\tau)>10$ GeV and $|\eta(\tau)|<2.0$ are selected; (3) For
other hadronic jets only those satisfying $P_T(jet)>60$ GeV and
$|\eta(jet)|<3.0$ are selected; (4) We require a large amount of
missing transverse momentum, $P^{miss}_T>200$ GeV;  (5) There are at least two jets that satisfy
the $P_T$ and $\eta$ cuts. Variations on (4) and (5)
will be discussed later.

The SM background  was determined  through the
generation of events from QCD multi-jets from  light and heavy quark flavors,
 Drell-Yan, single $Z/W$ production along with
quarks and gluons ($Z$/$W$ + jets), and $ZZ$, $WZ$, $WW$ pair
production which give multi-lepton backgrounds. We have cross
checked our SM backgrounds and several other elements of our
analysis with simulations done by CMS \cite{CMSnotes1,CMSnotes2} and
the results of these checks are in good agreement. We have also
found good agreement with the SUSY signal analyses of
\cite{CMSnotes1,CMSnotes2} and the total background analysis of
\cite{Baer:2007ya} under similar cuts. In PGS4 jets are defined
through a cluster-based  algorithm which has a heavy flavor
tagging efficiency  based on the parameterizations of the CDF Run
2 tight/loose SECVTX tagger \cite{CDF2} and is a displaced
(secondary) vertex b-tagging algorithm which allows detection of b
quarks. The b-tagging efficiency enters as a product of two
polynomials each a separate function of $|\eta(jet)|$ and
$P_T(jet)$. The efficiency is maximized in the region $|\eta(jet)| <
1$ with maximal efficiency ${\epsilon}_b =(0.4,0.5)$ for  tight and
loose tags respectively, and falls off sharply for $|\eta(jet)| > 1$
with virtually zero efficiency out near $|\eta(jet)|=2$ and $P_T(jet)
\sim 160$ GeV.  The analysis of tau decays is  done using 
Tauola\cite{Tauola}(For further details regarding PGS4 see \cite{PGS4wiki}).

There are several recent works which discuss dark 
matter\cite{Djouadi:2005dz,rosz,Allanach:2007qk,Feldman:2007fq,Kneur:2008ur,Belyaev:2007xa,baren,Allanach:2008iq,Barger:2008qd}
and collider  physics
\cite{Allanach:2004ud,Nojiri:2005ph,Baltz:2006fm,Carena:2006dg,Baer:2008uu,btag,UP,SP,Altunkaynak:2008ry,Kane:2008gb,Bhattacharyya:2008zi}.  The analysis
presented here is very  focused  in that it connects LHC signatures directly to the possible origin of dark matter,
and here we consider two dominant mechanisms, i.e., stau coannihilation and annihilation on  the HB.
We do not go into the details  of  the relic density 
analysis which are standard. 
Rather we go directly to  a discussion of how the experimental data, in particular the LHC data,   can allow one
to decode the mechanism for the generation of dark matter. 

\emph{Decoding the mechanism for the generation of dark matter with
LHC data:} It is well known that in the stau \co region, the
neutralino is typically  Bino like while in the HB region there is a
significant Higgsino component.  In the analysis of
\cite{Feldman:2007zn} it was shown that the Stau-Co region is
constituted of a collection of mass hierarchical patterns,  where mSP5 is
the  dominant pattern (defined by the mass hierarchy:  $\na$ $<$
$\sta$  $<$ $\slr$  $<$ $\snl$). Similarly  the analysis of
\cite{Feldman:2007zn}
 shows that  the HB region is dominated by the chargino patterns, where the chargino $\tilde \chi_1^{\pm}$ 
  is the NLSP,
and were classified in \cite{Feldman:2007zn}
as mSP1-mSP4 (where mSP1 is the dominant pattern defined by the mass hierarchy:
$\na$   $<$ $\cha$  $<$ $\nb$   $<$ $\nc$).
The largest cross sections in the direct detection experiments
arise from a Chargino Wall (CW) \cite{Feldman:2007fq} constituted uniquely of mSP1. Along the Wall
the thermal annihilation cross
sections in the early universe
would have arisen mostly from $\na \na$  annihilations into $WW, ZZ, t\bar  t, b\bar b$ 
for the Higgsino like LSP.
Additionally there are
a significant number of cases where the annihilation is  dominated by the processes
$\na \na \to b \bar b ~[\sim (85 - 90)\%]$ and $\na \na \to \tau^{+} \tau^{-}~[\sim (5 -10)\%]$, and
these cases typically occur
for larger neutralino masses which are more Bino like, but can also occur for low values of the neutralino
masses as well on the CW.

\begin{figure*}[t]
\includegraphics[width=7 cm,height= 6cm]{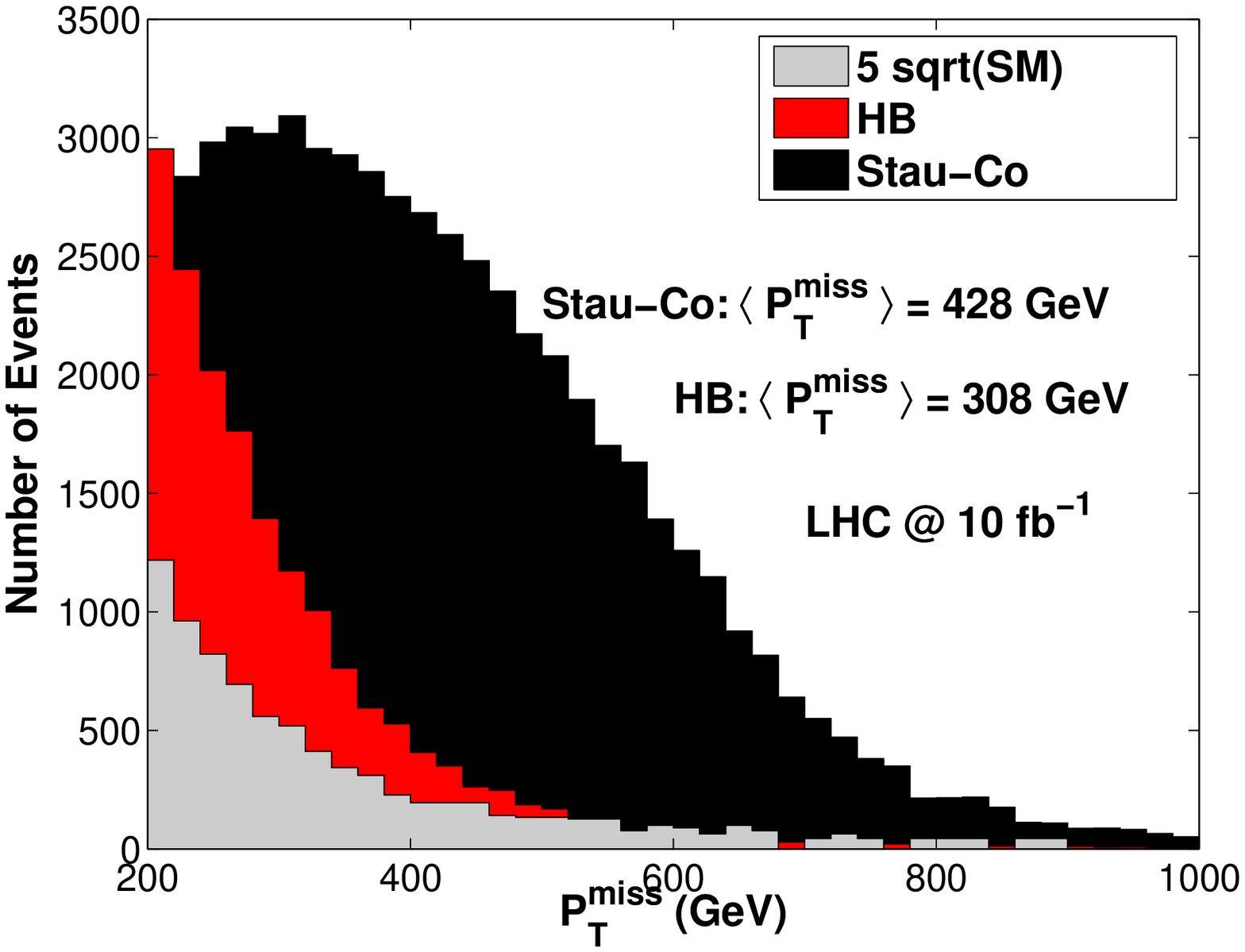}
\includegraphics[width=7 cm,height= 6cm]{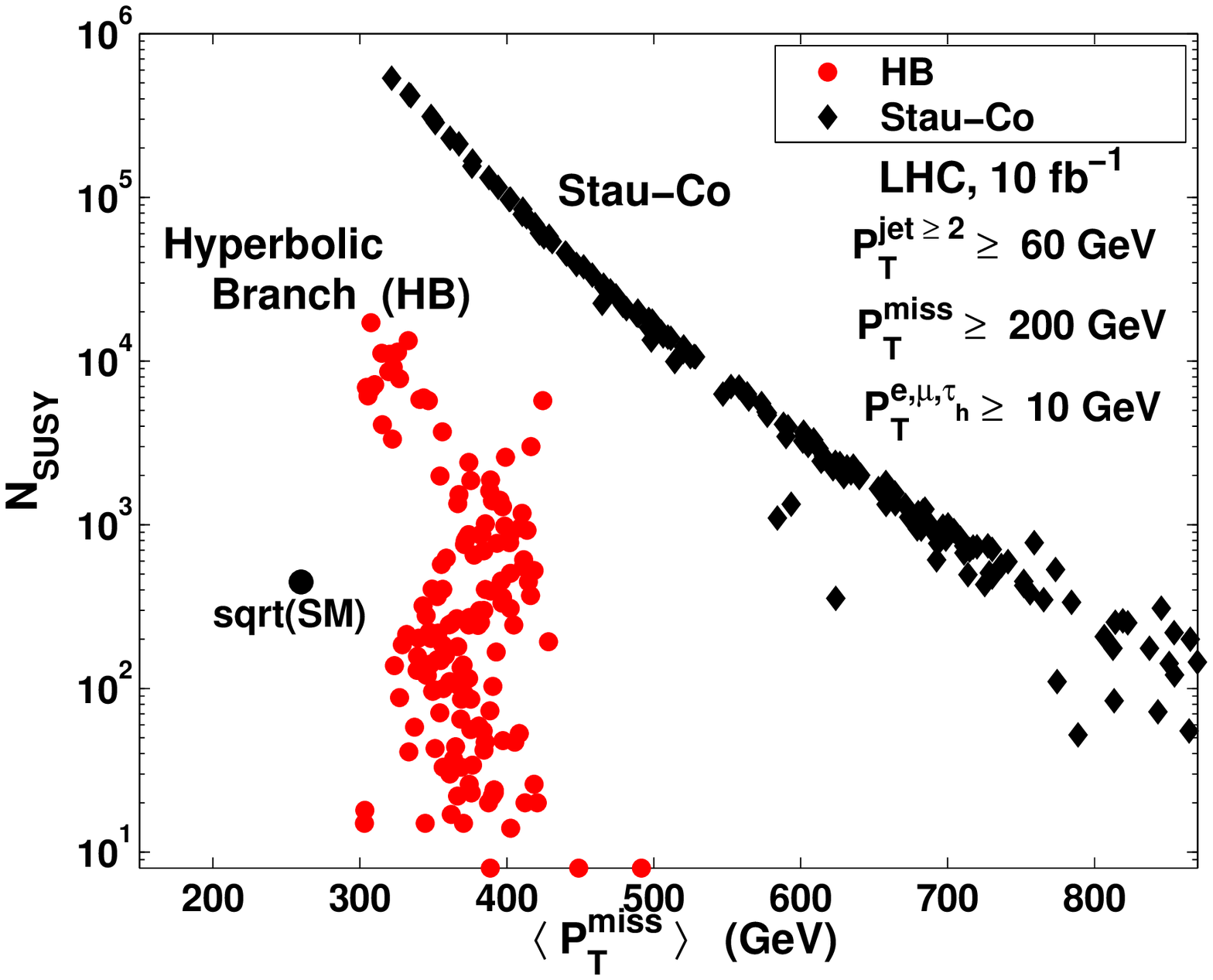}
\caption {
Left panel:
$N_{\rm SUSY}$ vs. \pt~ for  points in the Stau-Co and HB regions along with the SM background under 
the standard  post-trigger level cuts.
 The Stau-Co
and  HB  model points given here have
$(m_0,m_{1/2},A_0,\tan \beta, sign(\mu))$  as $(71.5,348,334,10,+)$ and
$(1694,216,-740,50,+)$ respectively (all masses are in GeV). The top mass is taken to be
170.9 GeV.
Right panel:
$N_{\rm SUSY}$ vs.  $ \langle
P^{\rm miss}_T \rangle$ for each parameter point in the Stau-Co and
HB.
 $ \langle
P^{\rm miss}_T \rangle$  acts as an indicator of  Stau-Co and HB regions.
 }\label{ptmiss}
\end{figure*}

The Chargino Wall referred to above enters importantly in the analysis of LHC signatures. 
It refers to the region of HB where the NLSP is the lightest chargino, the
LSP is mostly higgsino like, and the spin independent 
cross section is essentially constant $O(10^{-8})$ pb as function of the neutralino mass
for neutralino mass in the range  $\sim$(80-650) GeV. 
To explain this  feature we begin by exhibiting the LSP $\tilde \chi_1^0$  
state  in terms of the  gaugino-Higgsino  states so that 
$\na \equiv \chi =
 n_1{\tilde B}+ n_2{\tilde W}^3+n_3{\tilde H}^0_1 +n_4{\tilde
 H}^0_2.$
 The  LSP gaugino-Higgsino content enters
importantly in the thermal annihilation cross sections  that determine the
 proper relic density of decoupled neutralinos from the epoch of
freeze out.  
It also enters prominently in the strength of the scalar neutralino-proton  cross section 
which we now discuss.  
 Thus  
on the Wall the heavier of the CP Higgs mass has a lower limit
near 300 GeV and more typically it extends into the range of a TeV to several TeV.
On the CW  one typically has  $m^2_{H}
\gg m^2_h $,  and   $\sin \alpha \approx \alpha$ where
$\alpha$ is the Higgs mixing parameter which enters in the diagonalization of  the CP
even Higgs  mass$^2$ matrix. Further, the  sfermion poles can be neglected as they make
a small contribution in this region.
Under the above limits one finds that the product $\alpha \times \tan \beta$ is essentially
a constant, i.e., $\alpha \times \tan\beta \simeq -1$. Under this circumstance the
spin independent  cross section in the absence of CP phases is given by
\cite{Chattopadhyay:1998wb,Ellis:2000ds,Ellis:2008hf}
\beqn \sigma^{\rm SI}_{\chi p}({\rm WALL}) \sim \frac{m_p^2\mu^2_{\chi p}
g^2_2}{324 \pi  m^4_h M^2_W} (g_Y n_1 -g_2 n_2)^2\nonumber\\
\times(n_4+\alpha n_3)^2
(9 f_p+2  f_{pG})^2.
\eeqn
Here $\mu_{\chi p}$ is the reduced mass, 
and $f_p$ and $f_{pG}$ are  matrix elements defined by 
 $f_p=\sum_{i=u,d,s} f_i^p$, $f_{pG}= (1-f_p)$  where 
$m_p f_{i}^{(p)} =\langle p| m_{q_i} \bar q_i q_i |p\rangle$.
The typical ranges for
$n_i$ on the wall are:  $n_1
\in (.85,.99)$, $n_2 \ll n_1$, and $n_3 \in (.1,.6)\sim
-{\cal O}(n_4) $. Using  numerical values of $f_p, f_{pG}$\cite{Chattopadhyay:1998wb,Ellis:2000ds,Ellis:2008hf} 
 one gets
 $\sigma^{\rm SI}_{\chi p}({\rm WALL}) \sim 2  \times 10^{-8} $[pb].
  In our analysis presented later (see Fig.(5)),  however, we have implemented the full cross section calculation without
any of the above approximations. 
This analysis  leads to $\sigma^{\rm SI}_{\chi p}$ $({\rm WALL})$
lying in the range $\sim (1.5-5)\times 10^{-8}$ pb while the most recent limits
give $\sigma^{\rm SI}_{\chi p}\sim 5 \times 10^{-8}$ pb\cite{Ahmed:2008eu} for
$m_{\chi} \approx  60$ GeV. Thus this region of the parameter space is within  reach of the current and 
the next generation of dark matter experiments.  As noted already, the CW  is also a very interesting region
for LHC signatures. 
In the following we discuss several signatures  [listed as  (i)-(v)]  which  allow one to discriminate
 between the Stau-Co and the HB regions  over a significant part of the parameter space.

 (i) \emph{\pt Distributions:}
 A powerful signature for the discrimination of Stau-Co and HB  is the total number of SUSY events
  $N_{\rm SUSY}$  as a function of the
missing transverse momentum. An analysis of this signature is given in Fig.(\ref{ptmiss})
(left panel)
where one finds different geometries in these distributions.
Here we emphasize that the significantly fatter \pt distributions for points in the Stau-Co region  contrast sharply with the
\pt distributions from points in the HB region which are much thinner\cite{Feldman:2008hs}
as  exhibited for the case given  in Fig.(1)(left panel).
Such a signature has the interesting
feature in  that the discovery potential is increased over a larger
region in the SUGRA parameter space
 than for the case of counting fractional number of such events in separate channels.
 The above is due in part because every SUSY event that
passes the trigger has $ P^{\rm miss}_T$, and so one maximizes the
signal events as opposed to obtaining a fraction of them. Further,
the {\rm SM}  $ P^{\rm miss}_T$ falls off rapidly beyond the
peak value coupled with the fact the $ P^{\rm miss}_T$  is
larger in SUSY extending out to momenta where the {\rm SM} cannot
produce a large number of events.\\
(ii) \emph{$\langle P_T^{\rm miss}\rangle$ Analysis:}  A remarkable
signature emerges distinguishing the stau \co region  and the HB
region if one analyzes  $N_{\rm SUSY}$ for each parameter point as a
function of $\langle P_T^{\rm miss}\rangle$  which is the mean \pt ~calculated
by averaging the \pt ~over the entire model event record. The above phenomenon  is
shown in Fig.(\ref{ptmiss}) (right panel). Here one finds that
$\langle P_T^{\rm miss}\rangle$  has a very wide range from 300 GeV to a
TeV or more for the stau \co region, while  $\langle
P_T^{\rm miss}\rangle$~ for the HB region lies  in a much narrower band
centered around 350 GeV - a phenomenon which originates for
parameter points on the Chargino Wall. 
 Thus the $\langle P_T^{\rm miss}\rangle$ ranges
 in Fig.(\ref{ptmiss})(right panel) can be viewed as one of the smoking
gun signatures which can discriminate between the two mechanisms using LHC data.

Although a quantitative analysis of $\langle P_T^{\rm miss}\rangle$  is rather complicated
since it involves many particles and depends in part on  post trigger level cuts,  one can
give  a qualitative picture of the disparity  between the  $ P_T^{\rm miss}$ on the Stau-Co and
the HB regions by analyzing the decay chains of sparticles  into their final products 
culminating into an odd number of LSPs  (per sparticle decay chain) and the SM particles.
Here one finds that often the sparticle decays on the Stau-Co  involve two body decays.
For the HB case, however, one finds that  the sparticles produced in $pp$ collisions have typically
a longer decay chain  which depletes the $ P_T^{\rm miss}$ in this case.

We illustrate these features by
  analyzing the two specific benchmarks given in the caption of Fig.(\ref{ptmiss})(left panel).
For the HB model point of Fig.(\ref{ptmiss})(left panel) 
 the following production cross sections are
dominant: $pp \to (\g \g /  \nb \cha / {\widetilde{\chi}^{\pm}_1} {\widetilde{\chi}^{\mp}_1}$) at the level of
$(45,25,15)\%$.
While squark production is highly suppressed ($m_{\g} \sim 622 ~{\rm GeV} \ll m_{\ql,\qr,\ba, \ta} \in (1.2, 1.7)~{\rm TeV}$).
One finds that the dominantly produced $\tilde g$ decays via 
 ${\rm Br}[\g \to {\widetilde\chi}^{0}_{i} + q + \bar q]
\sim 50 \%$ and ${\rm Br}[\g \to {\widetilde\chi}^{\pm}_{j} + q + \bar q']
\sim 50 \%$ with the LSP contributing only $10 \%$.
The reason for  this largeness is because the on-shell decay of the gluino into $q\tilde q$ is suppressed due
to largeness of the squark masses (a phenomenon which typically holds for the gluino decays on the HB). 
Further, the $\nb$ and $\cha$ produced on HB have 
 ${\rm Br}[\nb \to {\widetilde\chi}^{0}_{1} + f + \bar f]
\sim 100 \%$ and 
 ${\rm Br}[{\widetilde\chi_1}^{\pm} \to {\na} +  f + \bar f']
\sim 100 \%$. Thus the decay chain for sparticles produced on the HB tend to  be longer resulting
in reduced $P_T^{miss}$
Now, for the Stau-Co model point  of Fig.(\ref{ptmiss})(left panel) 
 the leading SUSY
production level cross sections are from $(\g {\widetilde{q}},{\widetilde{q}}{\widetilde{q}}, \g\g)$ at the level of
$(41,33,7)\%$, with the corresponding 2 body decay modes ${\rm Br}[\qr \to \na + q]\sim 100\%$
(1st and 2nd generation), and   ${\rm Br}[\ql \to (\nb,\cha) + (q,q')]\sim (60,30)\%$.
Since the  decay chain for sparticles  on the Stau-Co tend to be shorter the resulting $P_T^{miss}$ is 
larger. 
In summary  the main reason why the HB tends to give lower values of missing $P_{T}$ relative the Stau-Co
is simply due to the fact that on the HB, in order to get to the LSP from the dominant gluino production
mechanism  one usually  needs at least 2 successive  3 body decays, while on the Stau-Co the right-squarks 
($\tilde q_R$) from the
first and second generations,
which are dominantly produced,  each decays right into the LSP + quark. The above also holds more 
 generally in that  
one finds that  sparticles arising from the Stau-Co have much shorter decay chains resulting 
in fewer final particles  and thus the missing energy  can get large.
Our more general results given  here on a large spread in $P^{\rm miss}_T$
 agree  with the first Ref. of \cite{Arnowitt} and  with the analyses of the CMS and ATLAS collaborations \cite{Yetkin:2007zz,Biglietti:2007mj,Migliaccio}.
Conversely the models on  the HB  have longer decay chains with more
final state particles and thus the missing energy carried by the neutrals is depleted leading to 
missing $P_T^{\rm miss}$ which is more SM like. 
This feature has also been discussed in  
the CMS analysis of \cite{mura}.

\begin{figure}[t]
\includegraphics[width=7.5 cm,height=5.0 cm]{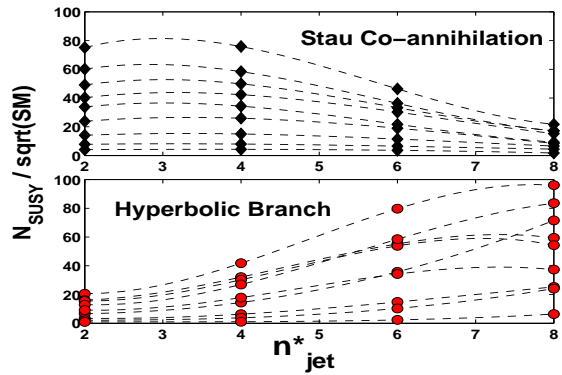}
\caption {A discrimination of the two mechanisms for the
satisfaction of the relic density in the early universe, i.e., stau
coannihilation and annihilation on  HB  with assumed LHC
luminosity of 10 fb$^{-1}$.   Curves
(connecting $N_{\rm SUSY}/\sqrt{\rm SM}$ for discrete $n_{jet}^*$)
 correspond to
 various models with $ M_{\rm LSP} <  275 ~{\rm GeV}$ for the Stau-Co
and $M_{\rm LSP} <  ~230~ {\rm GeV}$ for the HB.
The ratio of $N^{\rm HB/Stau-Co}_{\rm SUSY}/\sqrt{{\rm SM}}$ is computed
under the standard  post-trigger level cuts
(see also Fig.1)
but with $n^{*}_{jet}$ taken as a variable.
} \label{njet}
\end{figure}
\begin{figure}[t]
\includegraphics[width = 7.5 cm,height=5.0 cm]{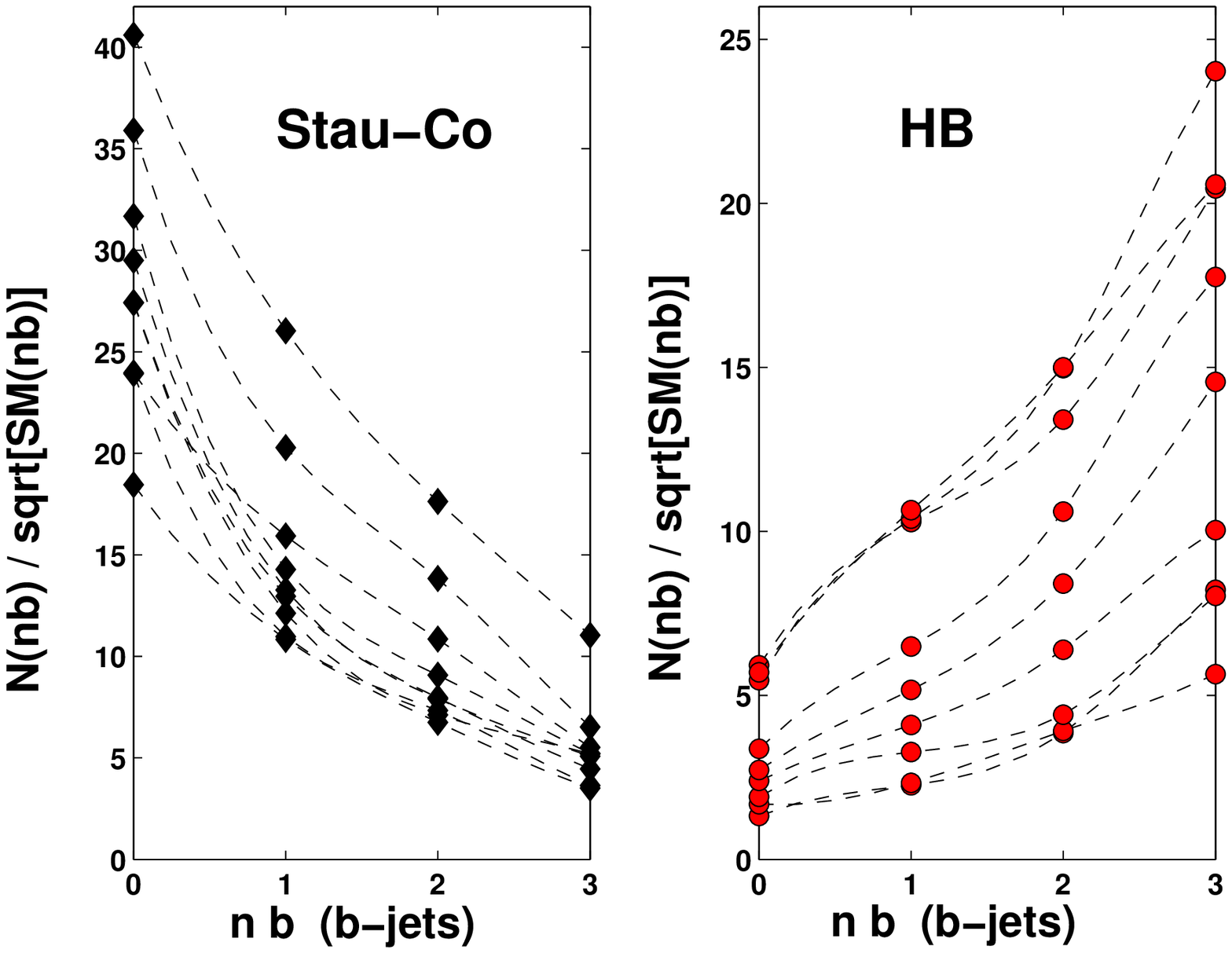} 
\includegraphics[width=6 cm,height=4.0 cm]{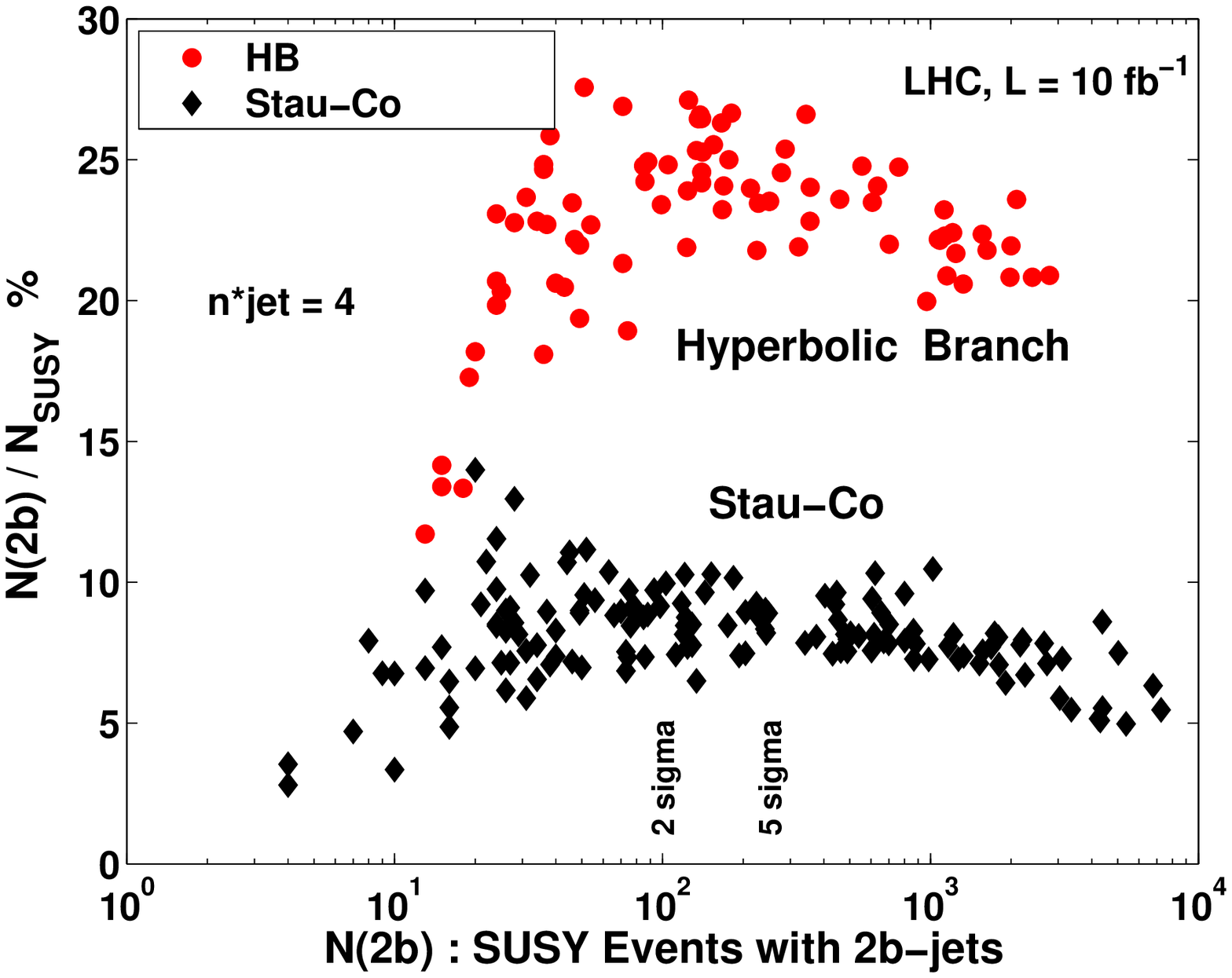}
\caption {
(a):  Top panels:  $N(nb)/\sqrt{{\rm SM}(nb)}$ vs  $nb$
for the Stau-Co and HB regions where $N(nb)$ (SM($nb$)) is the number of
SUSY (SM)  events that contain  $n$ b-tagged jets.
A sharp discrimination between the Stau-Co and the HB by b-tagging is observed.
The number $n^{*}_{jet}$ is fixed at $2$. Here $m_{\g} \leq 1.1$ TeV.
(b): Bottom panel:
A plot of  $N(2b)/N_{\rm SUSY}$  vs $N(2b)$  where  $n^{*}_{jet}$ is fixed at 4.
}
%
\label{fig:bjet}
\end{figure}

(iii) \emph{Jet cuts, \njs:}
Another powerful signature for discriminating the Stau-Co and the HB regions is
$N_{\rm SUSY}$ taken as a function of   \pts and \njs, i.e.,
\begin{equation}
N_{\rm SUSY}=N_{\rm SUSY}(n_{jet} \geqslant n_{jet}^{*}, P^{\rm miss}_{T}  \geqslant P^{*\rm miss}_T),
\label{sig}
\end{equation}
where the $*$ indicates a fixed cut value.
Here features specific to the  \co branch and to  the HB  emerge when
  $P^{*\rm miss}_T$ is fixed, and
 \njs~ is varied.
  This is shown in Fig.(\ref{njet}).
 In particular one finds that for the \co branch
there is an optimal $n_{jet}^*$ near 4  because the discovery limit
criteria $N^{\rm Stau-Co}_{\rm SUSY}/\sqrt{{\rm SM}}$
decreases as a function of increasing $n_{jet}^*$ and has a max near $n_{jet}$ $\sim$  4,
while for the
annihilations on the HB, specifically on the Wall, the situation is quite different,
in that the larger the $n_{jet}^*$ the larger is the value of $N^{\rm HB}_{\rm SUSY}/\sqrt{{\rm SM}}$
(where of course the {\rm SM} is subject to the  same * cuts).
Thus, as the jet number $n_{jet}^{*}$ becomes large   $N^{\rm HB}_{\rm SUSY}$
sustains a much stronger signal than $N^{\rm Stau-Co}_{\rm SUSY}$,
and thus $N_{\rm SUSY}/\sqrt{\rm SM}$ is a strong discriminator between the
Stau-Co and  the HB regions. 

The above becomes very significant if the SUSY scale is high with the LSP mass
lying in the several hundred GeV range. Also, this type of large n-jet
cut
can deplete the leptonic signal, so a delicate balance of jet cuts is very important. In fact,  if the
SUSY signal is not highly leptonic, the analysis of the above type would be an efficient way to
decipher new physics. Further,  even if one has signatures with many leptons, the jet analysis
will provide additional corroborating signatures for discovery and discrimination.
\\

\begin{figure}[t]
\includegraphics[width=7.5 cm,height=5.0 cm]{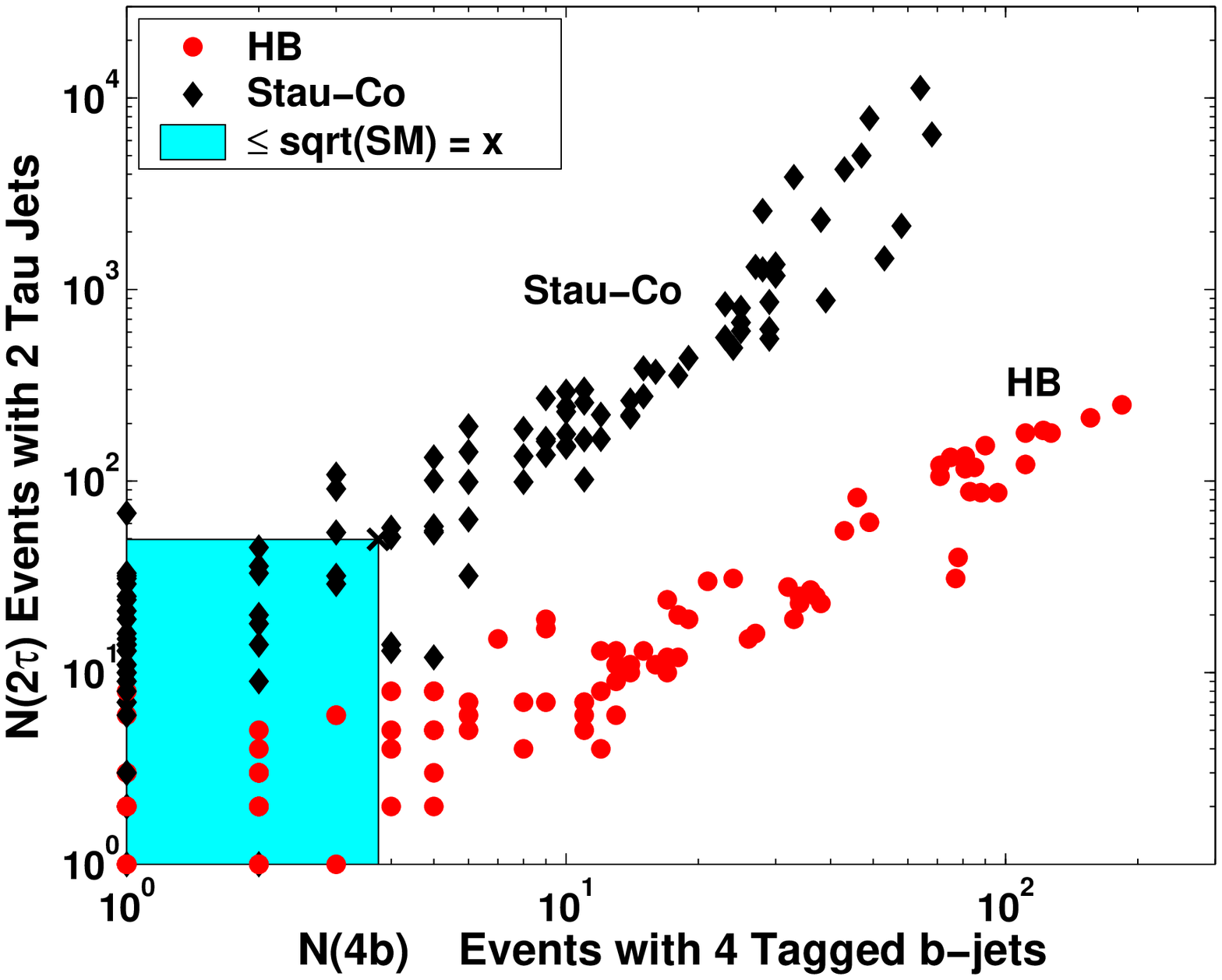}
\includegraphics[width=7.5 cm,height=5.0 cm]{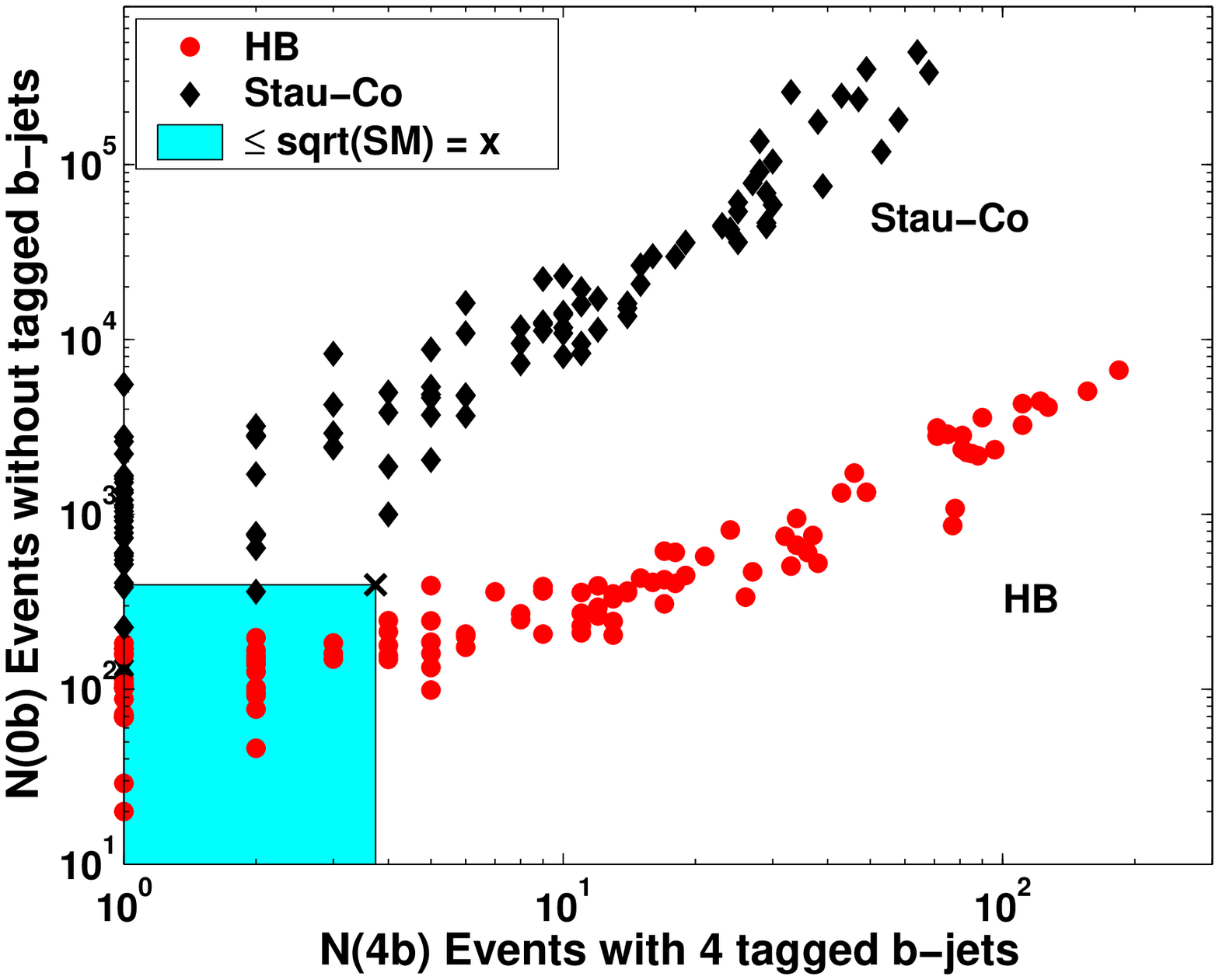}
\caption{
 $N(2\tau)$ (the number of events with two hadronically decaying $\tau$-jets) 
vs $N(4b)$ (the number of events with 4 tagged-b jets) (upper panel). A similar plot with
$N(2\tau)$ replaced by $N(0b)$ (the number of events with no tagged b-jets) (lower panel). 
The $\sqrt{\rm SM}$ values in each channel are indicated by 'X' on the plots.  Signatures arising from 
Stau-Co and from the HB regions are clearly discriminated. 
 } 
\label{correlate}
\end{figure}

(iv) \emph{Tagged  b-jets:}
The utility of b-tagging  for the HB region has previously been emphasized in
\cite{btag,Baer:2007ya,Feldman:2008hs}.
 In Fig.~(\ref{fig:bjet})(upper panels) we give an analysis exhibiting how b-tagging provides a striking
discrimination between the Stau-Co and HB regions where we plot $N(nb)/\sqrt{{\rm SM}(nb)}$ as a function of
the  number of tagged  b-jets  ($nb$) and find this dependence to be drastically different for Stau-Co vs HB
regions.   In Fig.~(\ref{fig:bjet})(lower panel) the fractional number of events with   2b-jets vs  
the number of events with  2b-jets  is given, and again one sees a strong discrimination between
the parameter points in the Stau-Co region vs those in the HB region.
  In Fig.(\ref{correlate}) we extend the analysis to the
4b-jet mode and correlate this signature to events with two hadronically decaying tau jets and to 
events that do not contain tagged b-jets. 

As already discussed in (ii),
on the HB 
 gluino production is dominant in $pp$ collisions at the LHC, 
 and further one finds  that the gluino decays dominantly into $b\bar b$, i.e., 
 ${\rm Br}[\g \to {\widetilde\chi}^{0}_{i} + b + \bar b]
\sim 40 \%$ and  ${\rm Br}[\g \to {\widetilde\chi}^{\pm}_{j} + b(\bar b) + \bar t (t)]
\sim 40 \%$.
Thus, the gluino  3 body decays are very rich in b quarks. 
Conversely  for the 
 Stau-Co model point of Fig.(1) (left panel)  the $pp$ production cross sections are as follows:
 $(\tilde g  \tilde q)\sim 41\%$,  $(\tilde q  \tilde q)\sim 33\%$,  $(\tilde g  \tilde g)\sim 7\%$ (as already noted
 in (ii)).
 Further, the gluino has only  a small branching ratio into $b\bar b$ in this case via 
 $\tilde g\to \tilde b b $.  Including the production cross sections for $\tilde g\tilde g$
 and the branching ratios, we find that overall the $b\bar b$ production on Stau-Co  
 is smaller relative to
 that on the HB. 
More generally the analysis of Figs.[(\ref{ptmiss})-(\ref{correlate})] shows that  
the LHC  signatures arising from the Stau-Co are easily distinguishable from 
those arising from the HB.

\begin{figure}[t]
\includegraphics[width=9.5 cm,height=6 cm]{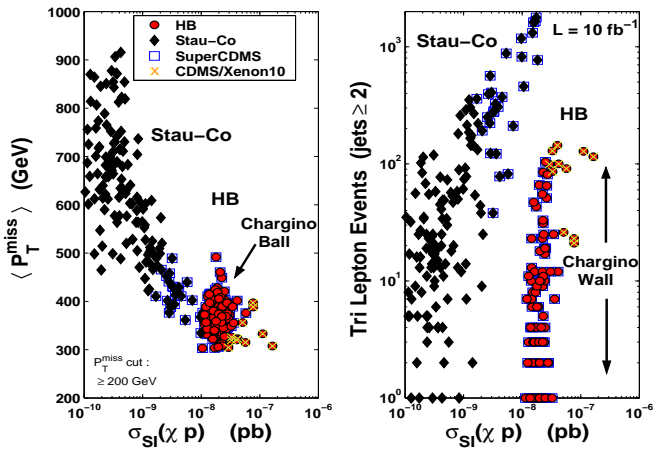}
\caption{Right panel: An exhibition  of the trileptonic signal  vs $\sigma^{\rm SI}_{\chi p}$.
Points in the vertical region to the right constitute the  Chargino Wall.
Left panel: an exhibition of $\langle P_T^{miss}\rangle$ vs $\sigma^{\rm SI}_{\chi p}$.
The cluster of points at the end to the right constitute the Chargino Ball.
The CDMS/Xe10 constraints\cite{Ahmed:2008eu} and  constraints expected 
from SuperCDMS\cite{Schnee:2005pj} are also shown.
A clear discrimination of Stau-Co and HB can be seen in these plots.}
\label{sigma}
\end{figure}

(v) \emph{$N_{\rm SUSY}(leptons/jets) ~{\rm and}~\langle P_T^{miss}\rangle
 ~{\rm vs}~ \sigma^{\rm SI}_{\chi p}$:}
Significant additional information regarding the \co region and the HB region
can be obtained 
 by an analysis of  the number of SUSY events at  the LHC  vs
the spin-independent neutralino-proton cross section $\sigma^{\rm SI}_{\chi p}$
 along with the current limits from the
direct detection of dark matter.  We give an illustration of the above in Fig.(\ref{sigma}).
The analysis  of Fig.(\ref{sigma}) (right panel) shows that the \co and the HB
 regions are well separated in the space spanned
by the trileptonic signature  3L (L=$e$,$\mu$)
and $\sigma^{\rm SI}_{\chi p}$. 
Further one observes that in Fig.(\ref{sigma}) (left panel) the parameter  points in the HB region
in the $ \langle P^{miss}_T \rangle - \sigma^{\rm SI}_{\chi p}$ plot are clustered together in a ball
shaped region and well separated from points in the Stau-Co region which lie on  a slope again
providing a strong discrimination between the Stau-Co and the HB regions.

  In the above we have analyzed in detail how the LHC data can allow one to discover if the 
 mechanism for the origin of dark matter in the early universe arises in the HB region or in the 
 Stau-Co region. This opens up the issue if the above  type analysis can be done more generally
 to identify the dominant mechanism for the generation of dark matter in the early universe 
 using LHC data. Below we discuss briefly how one may extend the analysis to 
 the stop coannihilation region and the A-pole  region. Thus  as already discussed in the introduction,
in addition to the Stau-Co region there is also a stop coannihilation (Stop-Co) region where the relic density constraint is
satisfied.   However, regions of the parameter space which
give rise to Stop-Co  have signatures which are  highly non-leptonic relative to those of
the Stau-Co and of the HB \cite{Feldman:2007zn,Feldman:2008hs}.
Thus correlations such as 0L + jets  vs 1L+ jets allow one to distinguish Stop-Co regions from 
others as discussed in the analysis of \cite{Feldman:2007zn,Feldman:2008hs}.
Such correlations 
 if observed,  would be a  good indication of stop coannihilation as the origin of dark matter.
 We note in passing that the LSP is mostly a Bino in this case which suppresses the scalar cross sections in the direct
detection of dark matter\cite{Feldman:2007fq}. 
Next  we discuss the $A$-funnel region.
 Here the relic density is satisfied  because the LSP mass is nearly half  the CP odd Higgs mass.
  The  analysis of parameter points which cluster near the pole region do not
 have the same NLSP for all parameter points unlike the case of HB where $\tilde \chi_1^{\pm}$   is the NLSP,
or  the Stau-Co region where $\tilde \tau_1$  is the NLSP.
 However, in the pole region   the NLSP could be $\tilde\chi_1^{\pm}$ or $\tilde \tau_1$ ($\tilde  t_1$ is
 seldom seen in the pole region).
Thus the $A$-pole region can give mixed  signatures, sometimes characteristic of  HB and
sometimes characteristic of the Stau-Co.  To firmly establish the pole region one would need a
 global analysis with many signatures which would give a determination
of the Higgs $A^0$ mass and the mass of the LSP.  A similar situation holds for other 
isolated regions of the parameter space which cannot be classified in the above categories
which satisfy the relic density constraints.  Here also one would need a global analysis on the 
signatures to identify the mechanism regarding the origin of dark matter.

\emph{Conclusions:} 
 The analysis  presented here shows that with
sufficient LHC data one can discriminate between the Stau-Co and the HB regions
regarding the origin of dark
matter in the early universe in the mSUGRA model using lepton, jet and missing
energy signatures. We discussed  several smoking gun signatures for 
such a discrimination. 
It was also shown that further discrimination is possible by combining
LHC data with the limits on $\sigma^{\rm SI}_{\chi p}$ from the direct detection of dark matter. 
An analysis in a similar spirit is given in \cite{Baltz:2006fm}.
Our analysis 
 utilizes simulations based on CMS detector specifications and specific final state signatures
 in probing the origin of dark matter,
 and we do so over a large portion of the parameter space. 
To our knowledge there are no analyses in the literature which have  tried to utilize the expected
LHC signatures to  probe the origin of dark matter along the lines discussed here.
While the analysis presented in our work 
illustrates our main points in the mSUGRA model, similar analyses along these lines should
be pursued for other models of soft breaking including string and  D brane models.
Specifically, it would be interesting to analyze what the LHC can tell us about the origin of dark matter
in cases where one has departures from mSUGRA including the case of a non-thermal relic
such as a Wino dominated neutralino.

{\it Acknowledgements}:
This work is supported in part by the U.S. NSF grant   PHY-075959.


\begin{thebibliography}{999}

\bibitem{msugra}
A.~H.~Chamseddine, R.~Arnowitt and P.~Nath,
  Phys.\ Rev.\ Lett.\  {\bf 49} (1982) 970.
%

\bibitem{hlw}
 L. Hall, J. Lykken and S. Weinberg, Phys. Rev. {\bf D27}, 2359 (1983);
 P.~Nath, R.~Arnowitt and A.~H.~Chamseddine,
  Nucl.\ Phys.\  B {\bf 227}, 121 (1983).

\bibitem{review}
For a review  see, P. Nath, R. Arnowitt and A.H. Chamseddine, "Applied
N=1 supergravity", (World Scientific,
Singapore,1984);
  H.~P.~Nilles,
  Phys.\ Rept.\  {\bf 110}, 1 (1984);
   P.~Nath,
  arXiv:hep-ph/0307123.

\bibitem{Spergel:2006hy}
  D.~N.~Spergel {\it et al.}
  Astrophys.\ J.\ Suppl.\  {\bf 170}, 377 (2007).

\bibitem{hb}
  K.~L.~Chan, U.~Chattopadhyay and P.~Nath,
  Phys.\ Rev.\ D {\bf 58}, 096004 (1998);
   J.~L.~Feng, K.~T.~Matchev and T.~Moroi,
  Phys.\ Rev.\ Lett.\  {\bf 84}, 2322 (2000);
 H.~Baer, C.~Balazs, A.~Belyaev, T.~Krupovnickas and X.~Tata,
  JHEP {\bf 0306}, 054 (2003).
For a review see,
  A.~B.~Lahanas, N.~E.~Mavromatos and D.~V.~Nanopoulos,
  Int.\ J.\ Mod.\ Phys.\  D {\bf 12}, 1529 (2003).

\bibitem{dmdd}
  U.~Chattopadhyay, A.~Corsetti and P.~Nath,
  Phys.\ Rev.\  D {\bf 68}, 035005 (2003).

\bibitem{Griest:1990kh}
  K.~Griest and D.~Seckel,
  Phys.\ Rev.\  D {\bf 43}, 3191 (1991).

\bibitem{Drees:1992am}
  M.~Drees and M.~M.~Nojiri,
  Phys.\ Rev.\  D {\bf 47}, 376 (1993).

\bibitem{Mizuta:1992qp}
  S.~Mizuta and M.~Yamaguchi,
  Phys.\ Lett.\  B {\bf 298}, 120 (1993).

\bibitem{Ellis}
  J.~Edsjo and P.~Gondolo,
  Phys.\ Rev.\  D {\bf 56}, 1879 (1997);
  J.~R.~Ellis, T.~Falk and K.~A.~Olive,
  Phys.\ Lett.\  B {\bf 444} (1998) 367;
  J.~R.~Ellis, T.~Falk, K.~A.~Olive and M.~Srednicki,
  Astropart.\ Phys.\  {\bf 13}, 181 (2000).

\bibitem{Nihei:2002sc}
  T.~Nihei, L.~Roszkowski and R.~Ruiz de Austri,
  JHEP {\bf 0207}, 024 (2002).

\bibitem{Nath:1992ty}
  P.~Nath and R.~L.~Arnowitt,
  Phys.\ Rev.\ Lett.\  {\bf 70}, 3696 (1993);
  Phys.\ Lett.\  B {\bf 299}, 58 (1993)
  [Erratum-ibid.\  B {\bf 307}, 403 (1993)];
 J.~L.~Lopez, D.~V.~Nanopoulos and K.~j.~Yuan,
  Phys.\ Rev.\  D {\bf 48}, 2766 (1993);
 H.~Baer and M.~Brhlik,
  Phys.\ Rev.\  D {\bf 53}, 597 (1996);
 V.~D.~Barger and C.~Kao,
  Phys.\ Rev.\  D {\bf 57}, 3131 (1998).

\bibitem{Arnowitt}
 R.~L.~Arnowitt, B.~Dutta, T.~Kamon, N.~Kolev and D.~A.~Toback,
  Phys.\ Lett.\  B {\bf 639}, 46 (2006);
  R.~Arnowitt, B.~Dutta, A.~Gurrola, T.~Kamon, A.~Krislock and D.~Toback,
  arXiv:0802.2968 [hep-ph]

\bibitem{Feldman:2007zn}
  D.~Feldman, Z.~Liu and P.~Nath,
  Phys.\ Rev.\ Lett.\  {\bf 99}, 251802 (2007).

\bibitem{Baer:2007ya}
  H.~Baer, V.~Barger, G.~Shaughnessy, H.~Summy and L.~t.~Wang,
  Phys.\ Rev.\  D {\bf 75}, 095010 (2007).

\bibitem{Feldman:2008hs}
  D.~Feldman, Z.~Liu and P.~Nath,
  JHEP {\bf 0804}, 054 (2008)
  [arXiv:0802.4085 [hep-ph]];
  arXiv:0806.4683 [hep-ph].

\bibitem{Djouadi:2002ze}
  A.~Djouadi, J.~L.~Kneur and G.~Moultaka,
  Comput.\ Phys.\ Commun.\  {\bf 176}, 426 (2007).

\bibitem{Belanger:2007zz}
  G.~Belanger, F.~Boudjema, A.~Pukhov and A.~Semenov,
  Comput.\ Phys.\ Commun.\  {\bf 177}, 894 (2007);
  arXiv:0803.2360 [hep-ph].

\bibitem{SLHA}
  P.~Skands {\it et al.},
  JHEP {\bf 0407}, 036 (2004).

\bibitem{Sjostrand:2006za}
  T.~Sjostrand, S.~Mrenna and P.~Skands,
  JHEP {\bf 0605}, 026 (2006).

\bibitem{PGS4} PGS4,  J.~Conway {\it et al.}


\bibitem{Djouadi:2006be}
  A.~Djouadi, M.~Drees and J.~L.~Kneur,
  JHEP {\bf 0603}, 033 (2006).

\bibitem{CMSnotes1} M.~Chiorboli, M.~Galanti, A.~Tricomi, CERN-CMS-NOTE-2006-133;
 D.~J.~Mangeol, U.~Goerlach, CERN-CMS-NOTE-2006-096.

\bibitem{CMSnotes2} W.~ de~Boer et. al, CERN-CMS-NOTE-2006-113.

\bibitem{CDF2}
 D.~E.~Acosta {\it et al.}  [CDF Collaboration],
  Phys.\ Rev.\  D {\bf 71}, 052003 (2005).

\bibitem{Tauola}
  S.~Jadach, Z.~Was, R.~Decker and J.~H.~Kuhn,
  Comput.\ Phys.\ Commun.\  {\bf 76}, 361 (1993).
  
\bibitem{PGS4wiki} http://v1.jthaler.net/olympicswiki/doku.php

\bibitem{Djouadi:2005dz}
  A.~Djouadi, M.~Drees and J.~L.~Kneur,
  Phys.\ Lett.\  B {\bf 624}, 60 (2005).

 \bibitem{rosz}
  L.~Roszkowski, R.~Ruiz de Austri and R.~Trotta,
  JHEP {\bf 0707} (2007) 075;
R.~R.~de Austri, R.~Trotta and L.~Roszkowski,
  JHEP {\bf 0605}, 002 (2006).

\bibitem{Allanach:2007qk}
  B.~C.~Allanach, K.~Cranmer, C.~G.~Lester and A.~M.~Weber,
  JHEP {\bf 0708}, 023 (2007);
  B.~C.~Allanach, C.~G.~Lester and A.~M.~Weber,
  JHEP {\bf 0612}, 065 (2006);
 F.~Feroz, B.~C.~Allanach, M.~Hobson, S.~S.~AbdusSalam, R.~Trotta and A.~M.~Weber,
  arXiv:0807.4512 [hep-ph].

\bibitem{Feldman:2007fq}
  D.~Feldman, Z.~Liu and P.~Nath,
  Phys.\ Lett.\  B {\bf 662}, 190 (2008)
  [arXiv:0711.4591 [hep-ph]].

\bibitem{Belyaev:2007xa}
  A.~Belyaev, S.~Dar, I.~Gogoladze, A.~Mustafayev and Q.~Shafi,
  arXiv:0712.1049 [hep-ph].

\bibitem{baren}
     G.~Barenboim, P.~Paradisi, O.~Vives, E.~Lunghi and W.~Porod,
  JHEP {\bf 0804}, 079 (2008).

\bibitem{Allanach:2008iq}
  B.~C.~Allanach and D.~Hooper,
  arXiv:0806.1923 [hep-ph].

\bibitem{Barger:2008qd}
  V.~Barger, W.~Y.~Keung and G.~Shaughnessy,
  arXiv:0806.1962 [hep-ph].

\bibitem{Kneur:2008ur}
  J.~L.~Kneur and N.~Sahoury,
  arXiv:0808.0144 [hep-ph].

\bibitem{Allanach:2004ud}
  B.~C.~Allanach, G.~A.~Blair, S.~Kraml, H.~U.~Martyn, G.~Polesello, W.~Porod and P.~M.~Zerwas,
  arXiv:hep-ph/0403133.

\bibitem{Nojiri:2005ph}
  M.~M.~Nojiri, G.~Polesello and D.~R.~Tovey,
  JHEP {\bf 0603}, 063 (2006).

\bibitem{Baltz:2006fm}
  E.~A.~Baltz, M.~Battaglia, M.~E.~Peskin and T.~Wizansky,
  Phys.\ Rev.\  D {\bf 74}, 103521 (2006).

\bibitem{Carena:2006dg}
  M.~S.~Carena, D.~Hooper and P.~Skands,
  Phys.\ Rev.\ Lett.\  {\bf 97}, 051801 (2006).

\bibitem{Baer:2008uu}
  H.~Baer and X.~Tata,
  arXiv:0805.1905 [hep-ph].

\bibitem{btag}
  H.~Baer, C.~h.~Chen, F.~Paige and X.~Tata,
  Phys.\ Rev.\  D {\bf 52}, 2746 (1995);
     U.~Chattopadhyay, A.~Datta, A.~Datta, A.~Datta and D.~P.~Roy,
  Phys.\ Lett.\  B {\bf 493}, 127 (2000);
  P.~G.~Mercadante, J.~K.~Mizukoshi and X.~Tata,
  Phys.\ Rev.\  D {\bf 72}, 035009 (2005);
    R.~H.~K.~Kadala, P.~G.~Mercadante, J.~K.~Mizukoshi and X.~Tata,
  arXiv:0803.0001 [hep-ph].

\bibitem{UP}
  U.~Chattopadhyay, D.~Das, A.~Datta and S.~Poddar,
  Phys.\ Rev.\  D {\bf 76}, 055008 (2007).

\bibitem{SP}
  S.~P.~Martin,
  Phys.\ Rev.\  D {\bf 75}, 115005 (2007);
  Phys.\ Rev.\  D {\bf 76}, 095005 (2007);
  [arXiv:hep-ph/0703097];
  arXiv:0807.2820 [hep-ph].

\bibitem{Altunkaynak:2008ry}
  B.~Altunkaynak, M.~Holmes and B.~D.~Nelson,
  arXiv:0804.2899 [hep-ph].

\bibitem{Kane:2008gb}
  G.~Kane and S.~Watson,
  arXiv:0807.2244 [hep-ph].

\bibitem{Bhattacharyya:2008zi}
  N.~Bhattacharyya, A.~Datta and S.~Poddar,
  arXiv:0807.0278 [hep-ph];
N.~Bhattacharyya, A.~Datta and M.~Maity,
  arXiv:0807.0994 [hep-ph].

\bibitem{Chattopadhyay:1998wb}
  U.~Chattopadhyay, T.~Ibrahim and P.~Nath,
  Phys.\ Rev.\  D {\bf 60}, 063505 (1999)
  [arXiv:hep-ph/9811362];
  A.~Corsetti and P.~Nath,
  Phys.\ Rev.\  D {\bf 64}, 125010 (2001).

\bibitem{Ellis:2000ds}
  J.~R.~Ellis, A.~Ferstl and K.~A.~Olive,
  Phys.\ Lett.\  B {\bf 481}, 304 (2000).




\bibitem{Ellis:2008hf}
  J.~Ellis, K.~A.~Olive and C.~Savage,
  Phys.\ Rev.\  D {\bf 77}, 065026 (2008).


\bibitem{Ahmed:2008eu}
  Z.~Ahmed {\it et al.}  [CDMS Collaboration],
  arXiv:0802.3530 [astro-ph];
J.~Angle {\it et al.}  [XENON Collaboration],
  Phys.\ Rev.\ Lett.\  {\bf 100}, 021303 (2008).
   
\bibitem{Schnee:2005pj}
  R.~W.~Schnee {\it et al.}  [The SuperCDMS Collaboration],
  arXiv:astro-ph/0502435.
   

\bibitem{Yetkin:2007zz}
  T.~Yetkin and M.~Spiropulu  [CMS Collaboration],
  Acta Phys.\ Polon.\  B {\bf 38} (2007) 661.

\bibitem{Biglietti:2007mj}
  M.~Biglietti {\it et al.},
  CERN-ATL-PHYS-PUB-2007-004; ATL-COM-PHYS-2006-095;ATL-PHYS-CONF-2007-020;
  ATL-COM-PHYS-2007-078.

\bibitem{Migliaccio}
A.~Migliaccio [ATLAS Collaboration], Diploma Thesis.

\bibitem{mura}B.~Mura [CMS Collaboration],  Diploma Thesis.



\end{thebibliography}
\end{document}